# Security Education in Higher Education through AI-Powered Gamification

Bingjun Li, Christopher Buzaid, Weihao Qu*

Monmouth University, West Long Branch, NJ, USA

Email: s1363772@monmouth.edu, s1363246@monmouth.edu, wqu@monmouth.edu

**Abstract** Cybersecurity education is facing more challenges as AI-driven attacks are becoming increasingly realistic and difficult to detect. Traditional video-based cybersecurity training in higher education often suffers from both low engagement and limited effectiveness. This dilemma motivates educators to explore innovative approaches, such as AI-powered gamification, which can deliver engaging, meaningful, and personalized learning experiences. By presenting content in a more interactive and user-friendly way, these methods have the potential to significantly improve both learner engagement and educational outcomes. This paper explores AI-powered gamification in cybersecurity education through the development of several short, mobile-friendly games. These games cover a range of topics from password security to text and phone scam recognition, incorporate multiple gamification strategies, including quiz-based, narrative-based, and simulation-based designs, as well as interactive formats such as TikTok Mini-Games. We conducted a two-tiered evaluation with 59 college students (comprising 9 technical experts and 50 general users), and the results indicate the potential of AI-powered gamification to improve engagement and increase attention to cybersecurity topics in higher education.

**Index Terms**—Cybersecurity, Gamified Learning, Gamification, Higher Education, Phone Scam, TikTok

## I. INTRODUCTION

Security has become an essential part of higher education. Institutions with large volumes of sensitive information, including student records, research findings, and intellectual property are attractive to cyber attacks [1], [2]. In particular, the rapid advancement of AI technologies makes cyber threats such as phishing, ransomware, and data breaches more legitimate and poses greater risks to teaching, research, and administrative systems [3]. A broadly accepted approach in higher education institutes is Security Education.

Security education in higher education aims to raise awareness, enhance knowledge, and promote positive behavioral changes among students, faculty, and staff [4]. However, two persistent challenges limit its impacts: **engagement** and **effectiveness**. Traditional training methods such as Mimecast or KnowBe4, which heavily rely on video-based tutorials and periodic quizzes, often suffer from both low engagement and limited long-term impact. First, these awareness trainings are often perceived as monotonous and disconnected from real-life scenarios [5]; Second, research further indicates that even when such awareness modules are completed, they rarely lead to measurable improvement in users' ability to recognize or respond to phishing attacks [6]. As a result, many learners complete the required modules without truly

Published in Journal of Cybersecurity, Digital Forensics and Jurisprudence, 2025, Vol. 1, pp. 65-80. DOI: 10.65879/3070-5789.2025.01.07.

internalizing cybersecurity practices, ultimately weakening the overall effectiveness of traditional approaches.

This persistent gap between awareness and behavioral change underscores the need for more innovative and interactive approaches to security education. One promising approach is **Gamification**, which refers to the integration of game design elements into non-game contexts to enhance motivation and engagement [7]. Gamification has been widely applied in education to increase learners' participation and sense of achievement [8]. Empirical studies show that gamified learning environments can improve student motivation, concentration, and learning outcomes across disciplines [9] - [11]. In higher education, gamification transforms passive instruction into active participation by combining challenges, feedback, and reward mechanisms that stimulate curiosity and persistence. These characteristics make gamification particularly well suited to security education, where learners must continuously adapt to complex and evolving risks.

However, traditional gamified platforms are not without limitations. Many are designed as **lengthy**, desktop-based systems that require substantial time and technical resources. In real higher-education settings, such formats can dramatically reduce accessibility and hinder learner engagement, especially as students often face heavy time constraints and prefer mobile-friendly learning experiences. When gamified modules are too long or inflexible, learners may lose interest before reaching the intended learning outcomes.

To overcome these barriers, recent studies have increasingly focused on integrating artificial intelligence (AI) into gamified environments to create more adaptive and personalized learning experiences. AI-powered gamification enhances traditional gamified learning by providing short, interesting, and personalized game experiences for learners. Through data-driven personalization, AI can identify specific weaknesses, such as difficulty in recognizing phishing messages or understanding password protocols, and it can also generate customized micro-learning modules to address these issues. This approach aligns with the growing emphasis on short and mobile-friendly educational interventions, which easily fit into students' daily routines and provide continuous reinforcement. By tailoring both content and difficulty to each learner's needs, AI-powered gamification makes security education more engaging, efficient, and sustainable.

This paper explores how AI-powered gamification can advance cybersecurity education in higher education institutions. It focuses particularly on short, mobile-friendly learning experiences that combine accessibility with adaptive personalization. Drawing on recent research and theoretical frameworks, this study analyzes how integrating AI-driven gamified strategies can overcome the limitations of traditional awareness training, enhance learner engagement, and strengthen the cybersecurity culture on university campuses. Its contribution includes:

1) the development of several short, mobile-friendly, AI-powered games on various cybersecurity aspects; and

2) evaluation of this sort of games on engagement (acceptance) and educational effectiveness among college students.

This paper takes an initial step toward demonstrating that AI-powered gamification not only enriches the learning process but also provides a scalable and effective model for fostering long-term behavioral change in cybersecurity education within the higher education environment

## II. BACKGROUND

Games serve as powerful educational tools by providing immersive, goal-oriented, and feedback-rich experiences that foster active engagement and facilitate deeper learning. According to Gee, well-designed games inherently model effective learning environments because they enable players to learn through experience, experimentation, and reflection rather than passive instruction [12]. Studies also show that games foster motivation, curiosity, and problem-solving by situating players within dynamic contexts where they can explore, collaborate, and apply knowledge to meaningful challenges [13]. Building on this understanding of how games facilitate meaningful learning experiences, researchers have proposed two primary approaches to incorporate game elements into education: **Gamification** and **Game-Based Learning** (GBL).

### A. Gamification versus Game-based Learning (GBL)

Gamification refers to the application of game design elements, such as points, levels, badges, and leaderboards, within non-game contexts to encourage motivation and persistence [7]. Huang et al. define it as a behavioral intervention strategy that integrates fun, feedback, and reward systems to maintain the engagement and commitment of learners in educational settings [14].

In contrast, Game-Based Learning (GBL) employs complete games or interactive simulations as the primary medium of instruction. It focuses on experiential learning, problem-solving, and situated cognition, allowing learners to acquire knowledge and apply skills within authentic, interactive environments [15], [16].

Although both Gamification and Game-Based Learning aim to inspire engagement and behavioral change, they differ in scope and implementation [17]. Gamification improves existing learning activities by incorporating motivational game elements [7], [14]. On the other hand, GBL transforms the learning activity itself into a complete game experience [13], [15]. In other words, gamification influences how students learn by enriching traditional instruction, while GBL redefines what and where they learn through immersive gameplay. According to Caponetto et al., early research often blurred the distinction between these two concepts, but recent studies show that gamification serves as a motivation layer over conventional contexts, while GBL constitutes a fully integrated learning environment [18], [19].

### B. Current Challenges in Security Education

Cybersecurity education in higher education faces escalating challenges as digital threats evolve rapidly and become increasingly sophisticated [20]. Institutions have become prime targets for phishing, ransomware, and social engineering attacks, many of which exploit human behavior rather than purely technical vulnerabilities [21]. These realities underscore the urgent need for more adaptive, interactive, and learner-centered approaches that go beyond passive awareness modules [22]. Traditional approaches, such as Mimecast and KnowBe4, are often static and outdated. They are heavily based on repetitive video tutorials or standardized quizzes that fail to foster engagement or produce lasting behavioral change. As a result, many participants often complete mandatory modules without developing the situational awareness or critical thinking skills necessary to respond effectively to complex threats in the real-world [2].

### C. Case Study on the Limitation of Traditional Training

Fig. 1 : The Case Study of Real-world AI-powered Text Scam

The limitations of traditional cybersecurity education became especially clear through a recent real-world incident, which demonstrated the urgency of more adaptive and experiential learning approaches. As shown in Figure 1, in September 2025, a graduate student under investigation, who had completed the required security awareness training required by her university, received a text message that appeared to originate from an official U.S. government agency. The message stated that a tax refund claim had been processed and required immediate confirmation of payment information before a specific deadline. The message contained a link that closely resembled a legitimate government website, and its visual design, tone, and language conveyed an authentic sense of urgency and authority. At first glance, the message appeared entirely legitimate. Believing it to be authentic, the graduate student provided personal information as well as credit card details. Shortly afterward, the provided card was added to a different digital wallet, as illustrated in the right panel of Figure 1.

However, a closer inspection revealed subtle inconsistencies. The sender's phone number is from the Philippines, and the embedded URL redirects to a fraudulent domain designed to mimic the official site. This message was deliberately engineered to exploit psychological triggers, particularly urgency and credibility, leading to hesitation even among individuals familiar with technology and digital safety.

This incident illustrates the inadequacy of conventional awareness programs, which often rely on repetitive tutorials or standardized quizzes that fail to address the increasing sophistication of phishing attacks [23], especially those enhanced by generative artificial intelligence [24]. Nowadays, scammers can create context-sensitive personalized messages that closely mirror official communications, effectively bypassing both technical filters and human skepticism [25]. These static approaches cannot effectively foster the critical thinking or situational awareness required to recognize and respond to evolving digital threats.

As attackers increasingly utilize AI-generated automation and content, cybersecurity education must evolve accordingly [26]. The attack's effectiveness aligns with the findings of Ferreira et al. [27], who demonstrated that phishing campaigns rarely rely on a single trigger. Instead, they employ layered persuasion, most commonly pairing Authority with Distraction or Liking strategies to suppress critical thinking.

Future training should emphasize interactive, scenario-based, and experiential learning that mirrors authentic attack situations and engages learners in active decision-making [28]. Only through such adaptive and immersive approaches can educational institutions effectively prepare individuals to identify and counteract dynamic, AI-driven cyber threats.

### D. Gamification in (Cybersecurity) Education

The increasing sophistication of cyber threats demands educational approaches that actively engage learners in realistic, high-stakes scenarios rather than relying on passive information delivery. Gamification provides a promising framework for this transformation, turning cybersecurity instruction into an interactive and experiential learning process [29]. Rather than simply memorizing abstract principles, learners are immersed in simulated challenges that replicate real-world attacks, enabling them to apply knowledge, make decisions, and observe the consequences of their actions within a safe and controlled environment.

Recent research highlights the effectiveness of gamified cybersecurity education in enhancing learner engagement, knowledge retention, and practical performance [9] – [11], [30], [31]. Studies have shown that gamified cybersecurity education can significantly enhance both conceptual understanding and technical proficiency by fostering active participation, collaboration, and problem-solving in realistic learning environments [32]. For instance, Kim et al. [33] conducted a multi-study analysis and participated in gamified cybersecurity labs, demonstrating stronger motivation, higher engagement, and improved learning outcomes compared to those in traditional instructional settings. These findings highlight the transformative potential of gamified learning environments to enhance both engagement and performance in cybersecurity education.

These findings suggest that gamification can transform cybersecurity education from compliance-based instruction to an active, experiential learning process that strengthens both competence and confidence. However, most existing gamified systems still rely on fixed difficulty levels, limiting their ability to adapt to individual learners' progress and performance [34]. This limitation highlights the need for more adaptive, data-driven approaches that can personalize challenges and feedback according to learners' progress and abilities.

### E. AI-powered Gamification

To overcome these limitations, recent studies are increasingly exploring AI-powered gamification, which uses adaptive intelligence to personalize learning experiences and sustain engagement in real time [35]. Unlike traditional gamified systems, AI-powered platforms continuously analyze learners' behavior and performance in real time to dynamically adjust task difficulty, personalize learning pathways, and deliver targeted feedback. This adaptivity ensures that learners are appropriately challenged and engaged, promoting steady skill development and deeper understanding [36], [37].

In the context of cybersecurity education, such systems can simulate evolving threat scenarios that adjust to a learner's proficiency level, fostering situational awareness and decision-making under pressure [38]. As digital threats continue to grow in sophistication, AI-powered gamification represents a promising approach to developing more resilient, adaptive, and engaging cybersecurity education in higher education institutions [29].

## III. GAMES

### A. AI-powered Gamification in Education

The first game we introduce highlights the significant potential of AI-powered gamification to enhance learning in educational settings. It was developed by an undergraduate computer science student who initially had limited interest in his coding class. Instead, with the help of AI tools, he created a mobile-friendly game that helped him stay motivated, review key concepts, and prepare for exams, as shown in Figure 2. Figure 2 (a) shows the main entry of the game, while Figure 2(b) and (c) depict the gameplay experience, demonstrating how gamification can transform learning into an engaging and effective process. This game is available at: https://python-demo-eight.vercel.app/.

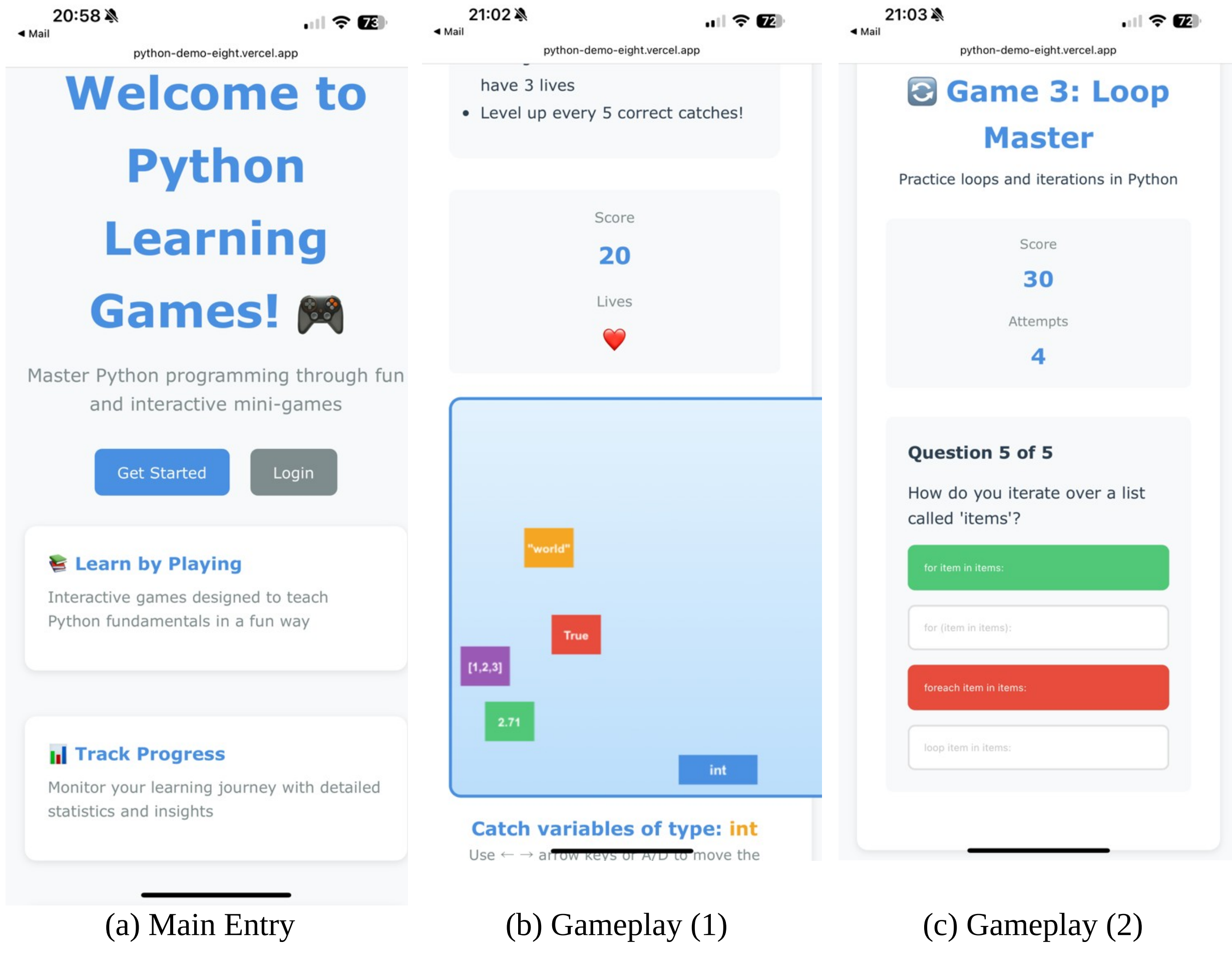

(a) Main Entry (b) Gameplay (1) (c) Gameplay (2)

Fig. 2: AI-powered Python Games Developed by a Student.

### B. Sentinel Security Game Platform Design

We developed an AI-powered game platform called **Sentinel**, which provides a collection of subgames covering essential security topics such as phishing detection, password cracking, text scam, and phone call scam awareness as shown in Figure 3(a). The implementation involves generative models and logic. The core interaction engine is built upon a Large Language Model (LLM) framework (specifically Google Gemini API) fine-tuned with system prompts that define the "persona" of the attacker.

**1) Persona Definition:** For the Phone Scam game, the system prompt restricts the AI to specific social engineering tactics (e.g., urgency, authority) and prevents it from breaking character.

**2) Voice Synthesis:** The textual output from the LLM is piped through a low-latency Text-to-Speech (TTS) engine (ElevenLabs) to generate realistic voice modulation, including pauses and intonation changes that mimic human conversation.

**Adaptation and Constraints:** To ensure educational safety, the model operates within a "Bounded Adversarial" framework. First, user inputs are filtered to prevent "jailbreaking" the scammer bot, such as convincing the bot to reveal it is an AI. Also, the game platform adjusts the sophistication of the scam based on the user's initial streak. If a user easily detects the first clue, the AI is prompted to switch to a more subtle "soft-sell" tactic in the subsequent turn.

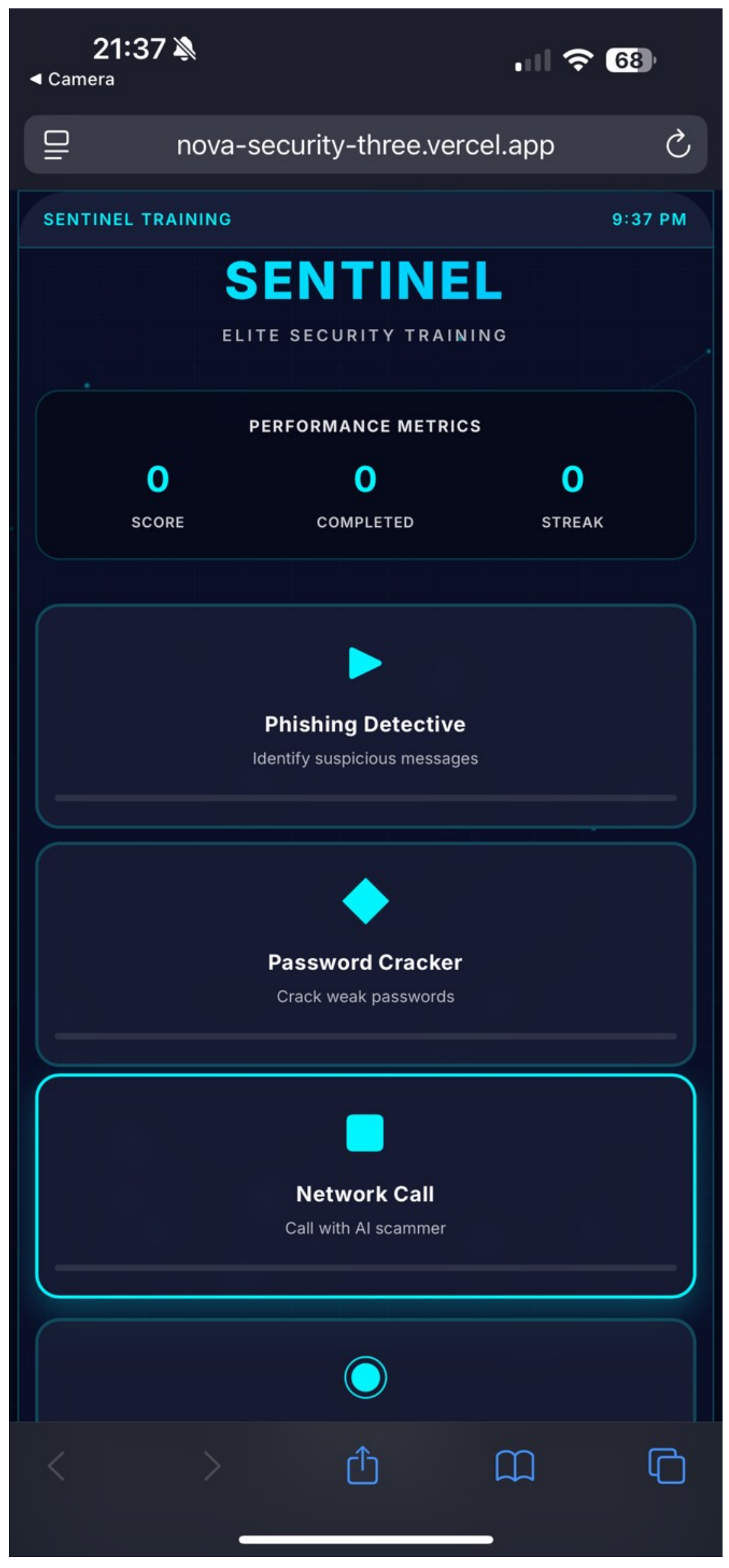

(a) Main Entry

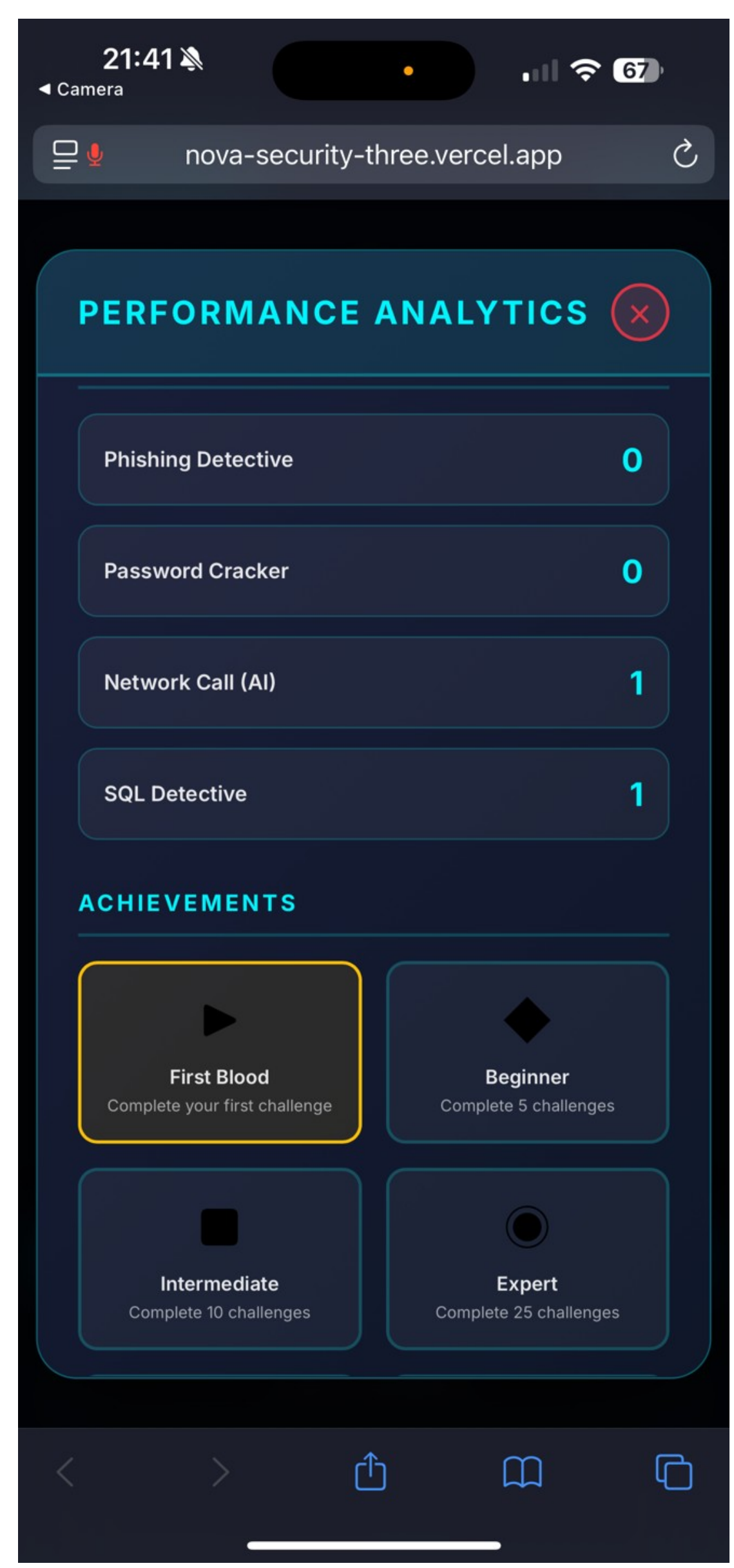

(b) Badge in Sentinel

Fig. 3 Sentinel Security Game Platform on a Mobile Phone.

### C. Mini games

Sentinel Game Platform provides player analytics and a badge-reward system to enhance motivation and sustained engagement. As shown in Figure 3(b), the platform tracks player performance and awards badges based on progress and achievements, reinforcing continued participation and learning. A series of mini games will be available at: https://www.cyberbeangames.com.

**a) The Phishing Detection Mini-game:** The phishing detection subgame simulates realistic phishing emails and raises learners' security awareness by challenging them to identify suspicious or malicious elements. For instance, in Figure 4(a), learners need to analyze a phishing email disguised as a Linkedln Job Opportunity and need to pinpoint the indicators of fraud.

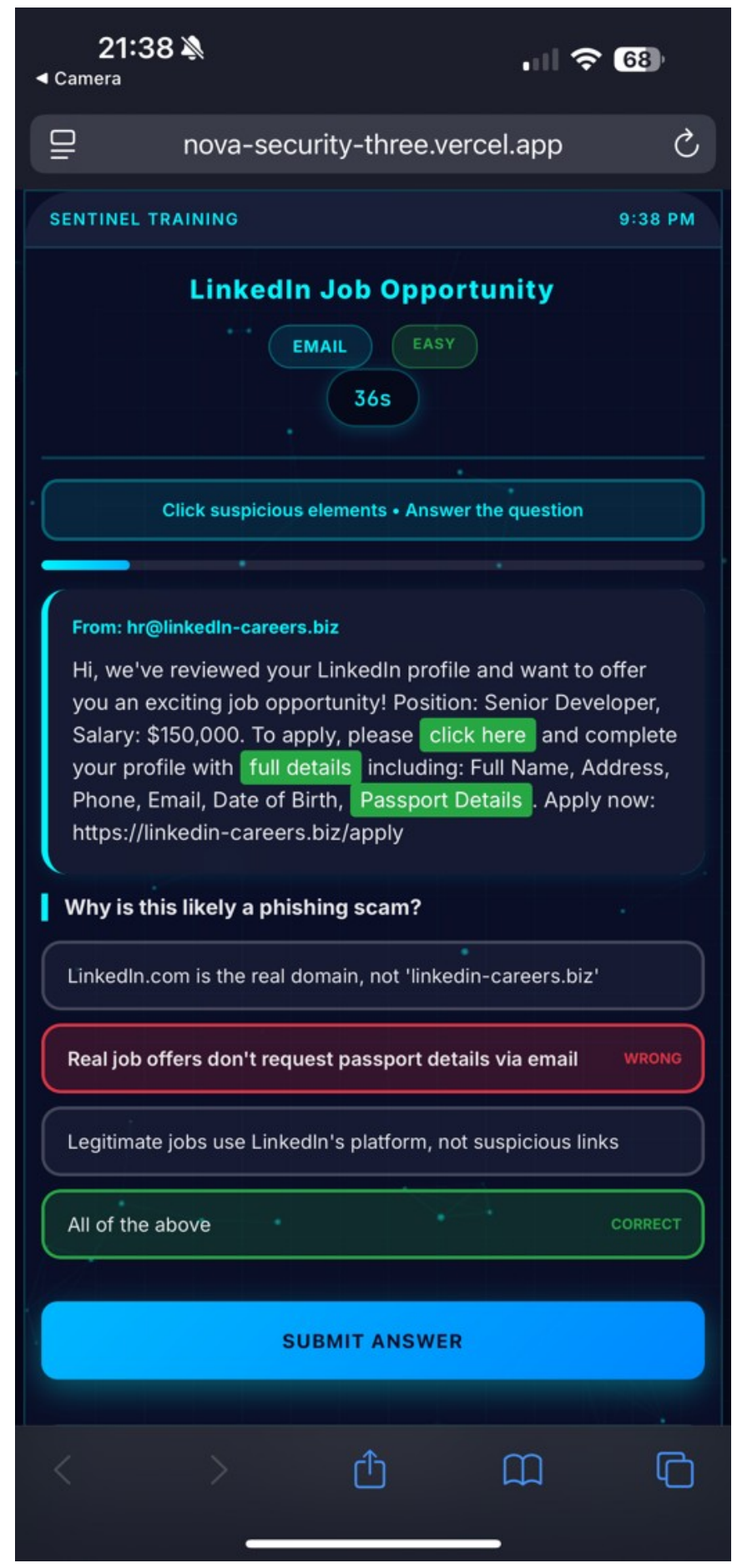

(a) Phishing Detection

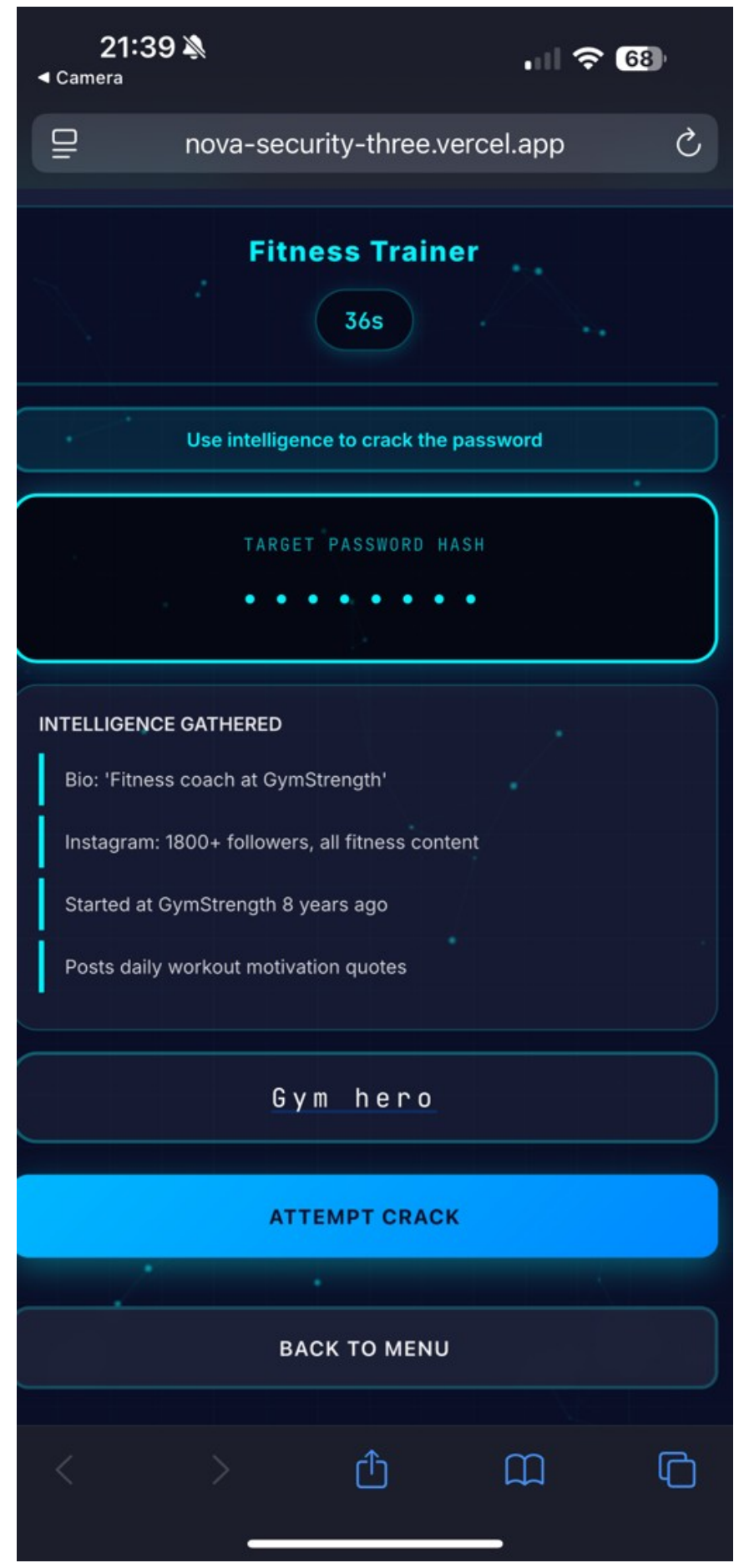

(b) Password Cracker

Fig. 4: AI-powered Security Game Platform Sentinel.

**b) The Password Cracker Mini-game:** As shown in Figure 4(b), this game provides players with brief background information about a fictional target and asks players to guess the password. By engaging in this activity, students can develop a deeper understanding of common password weaknesses, and they are encouraged to avoid similar mistakes in their own password choices.

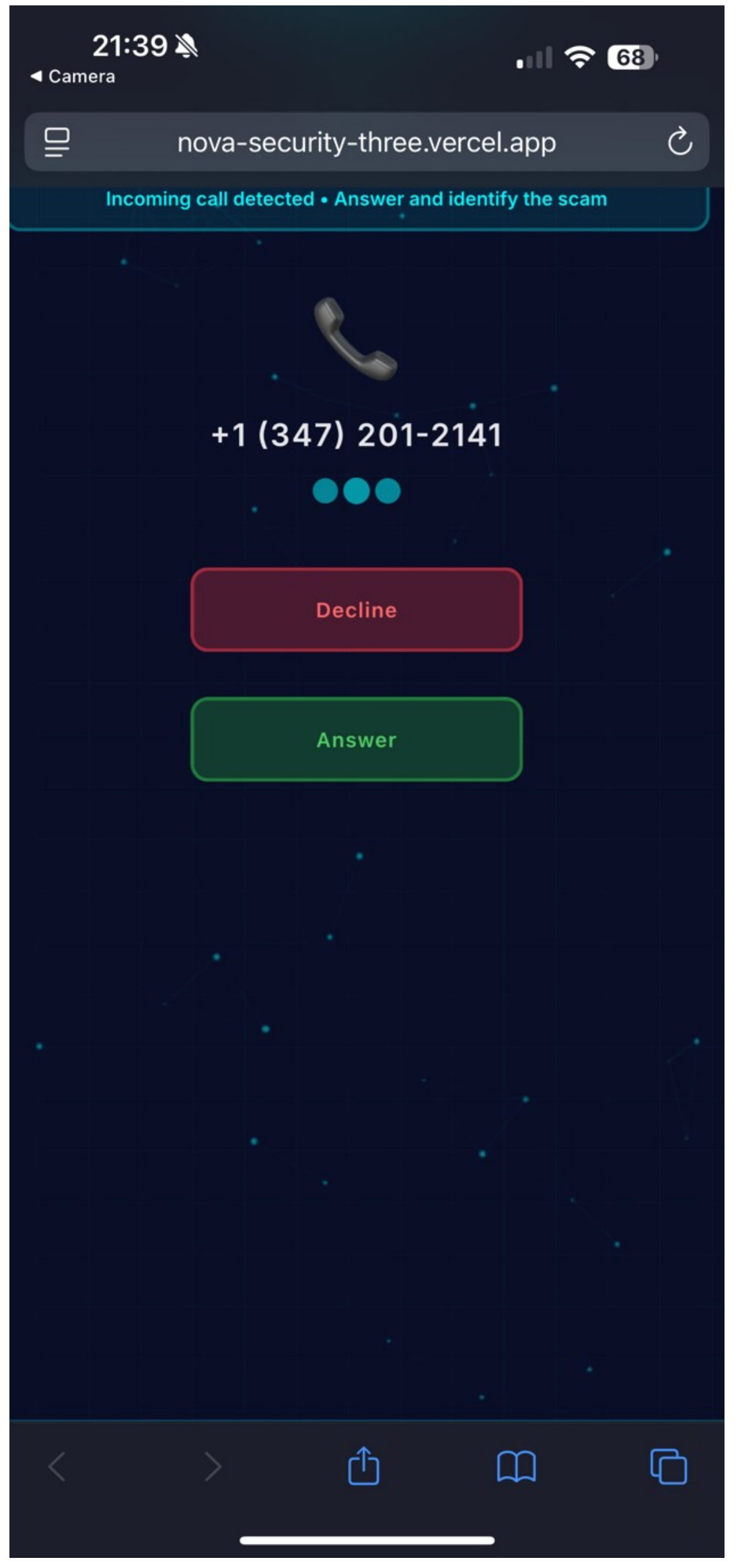

(a) Phone Simulation

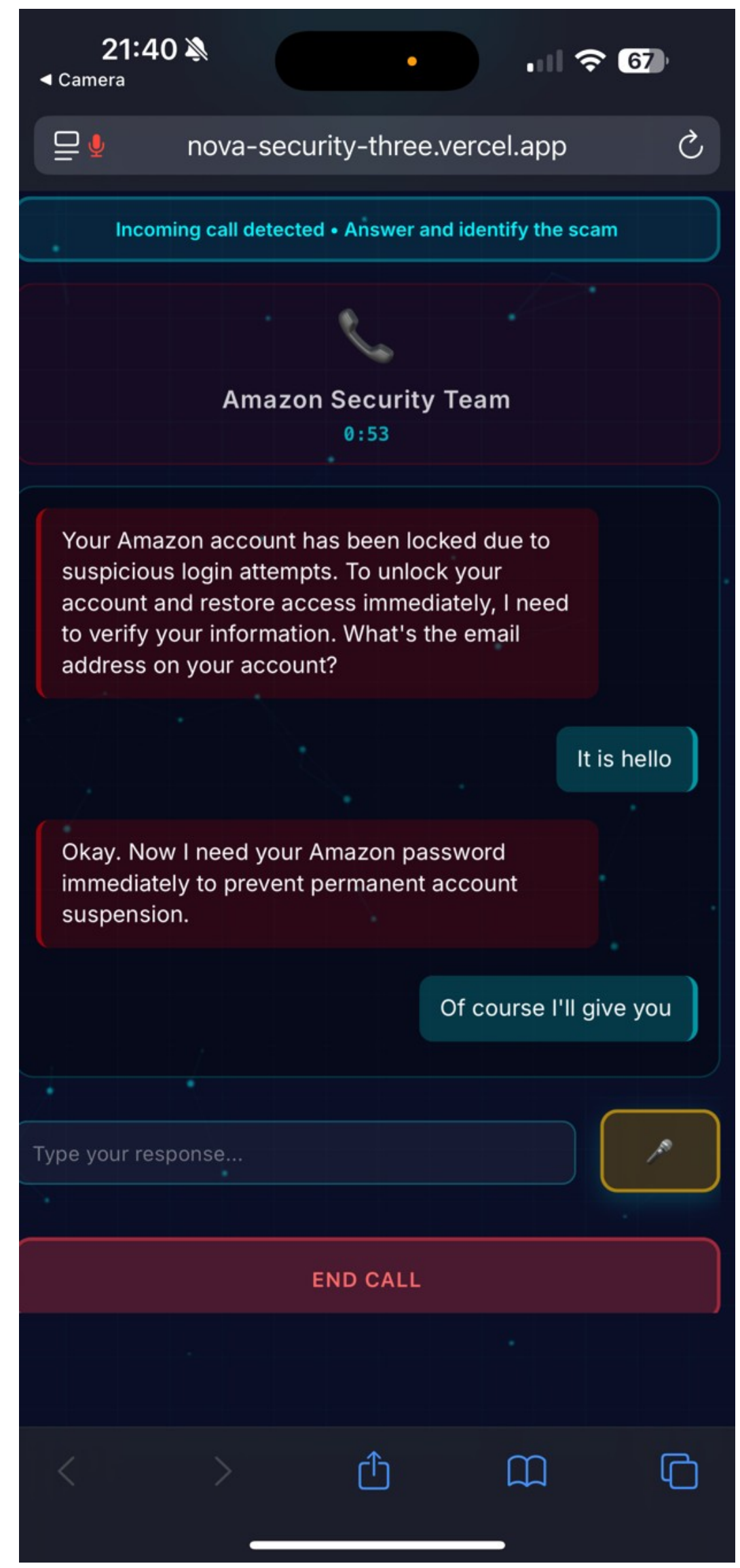

(b) Phone Scammer Dialogue

Fig. 5: Game Play of Sentinel Phone Scam Game.

**c) The Phone Scam Recognition Mini-game:** The phone scam recognition subgame focuses on social engineering threats. In this game, the user will receive a simulated phone call as shown in Figure 5(a). Our specific trained AI bot, shown in Figure 5(b), will act as a scammer and attempt to extract sensitive information, such as passwords or account details. This interactive scenario helps learners practice recognizing manipulation tactics and responding safely.

**d) TikTok Filter Mini-game (Password Thinker):** We also explored the potential of short-video-style mini-games for cybersecurity education. Figure 6 presents a TikTok Filter Game in which users are asked to choose a stronger password between two options. Players need to complete five rounds within a total of 20 seconds, which creates a fast-paced and engaging learning experience. This game represents our first attempt to use short-video filter formats to increase user engagement and deliver essential cybersecurity concepts in under 30 seconds.

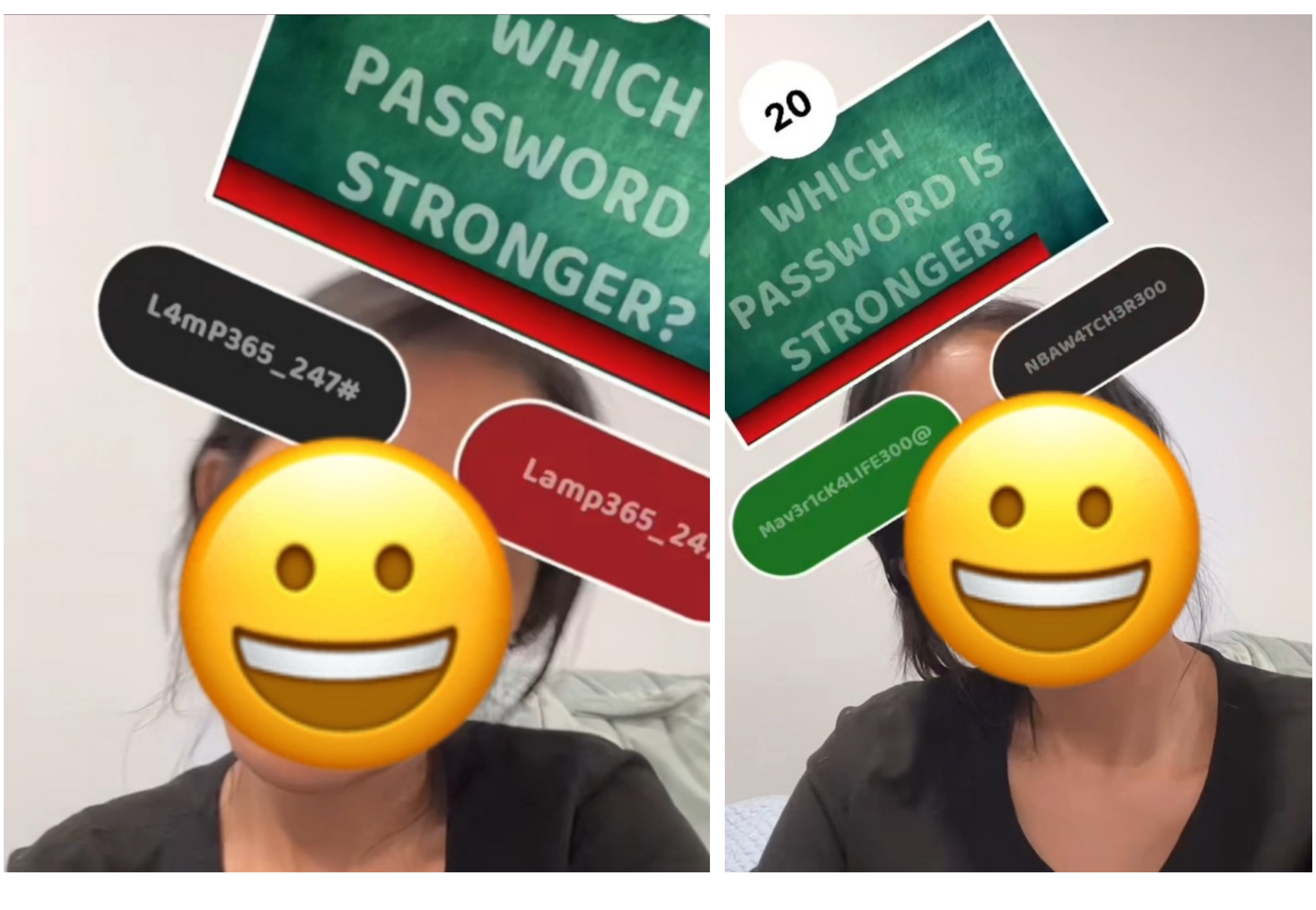

(a) Wrong Gameplay

(b) Correct Gameplay

Fig. 6: TikTok Filter Game: Password Thinker.

Above all, these short, mobile-friendly games illustrate how AI-powered gamification can make cybersecurity learning more active, practical, and engaging in higher education. Gamified security education provides learners with hands-on experience through interactive simulations and competitive problem-solving tasks [38]. Such approaches help bridge the gap between theoretical understanding and practical application by replicating real-world security scenarios that require timely decision-making and strategic thinking. Studies show that gamified environments can effectively model authentic cyber-defense settings and assess situational awareness under realistic stress conditions [39].

## IV. METHODOLOGY AND EVALUATION

To evaluate both technical robustness and broad pedagogical acceptance of the gamified mobile security education platform, we adopted a Two-Tiered evaluation strategy involving two distinct participant groups.

### A. Study Design and Participants

**a) Tier 1: Expert Review and Technical Validation (N=9):** The first group consisted of nine (N=9) Computer Science students enrolled in an advanced Cybersecurity course. As the developers and alpha testers of these games, this group possessed domain-specific knowledge. Their role was to evaluate the technical accuracy of the security scenarios, the realism of the AI responses, and the stability of the mobile deployment.

**b) Tier 2: General Learner Preference Survey (N=50):** To assess the wider acceptance of gamified, mobile-first security education and ensure a representative sample, we recruited an equal gender split of 25 males and 25 females (N=50). Crucially, the participants represented a diverse range of academic disciplines, ensuring that the results reflect a general user base rather than solely technology specialists. As shown in Figure 7, the largest single cohort was from the Education department (N=16). This is followed by Computer Science \& Math (12 students), which is heavily weighted towards Computer Science majors(10), and Data Science(1), Math(1). Two categories, Business \& Finance and Health \& Life Sciences, are tied with 9 students each, showing a diverse mix of related majors within them. The Business \& Finance group is led by Finance (4) and Business (3), with Marketing and Economics contributing one student each. The Health \& Life Sciences category is fairly evenly distributed among Psychology (3), Biology (2), Nursing (2), and Health Sciences (2). The smallest grouping is Social \& Legal Studies, comprising 4 students consisting of Criminal Justice (2), Homeland Security (1), and Anthropology (1).

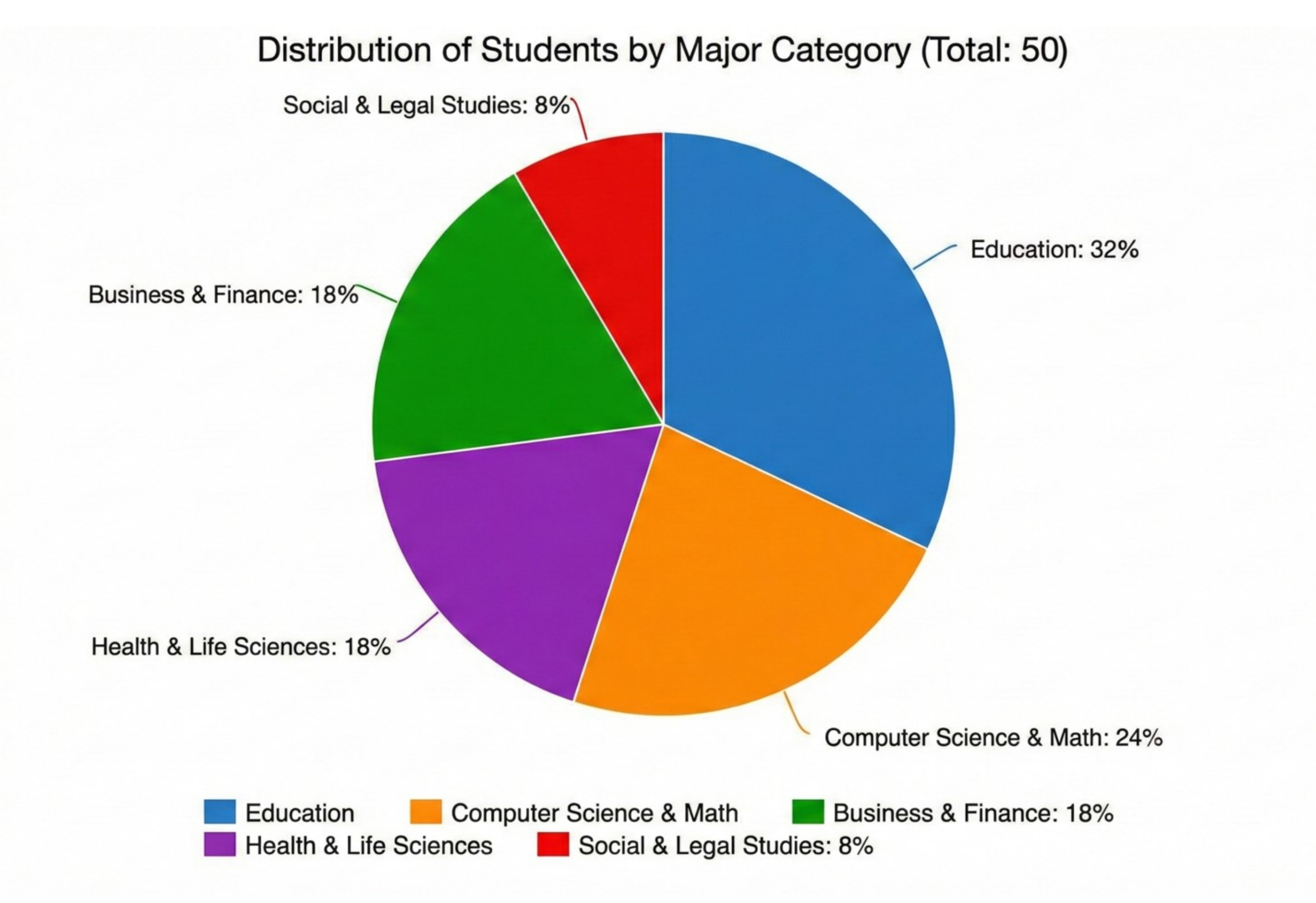

Fig. 7: The major distribution of participants in Group 2.

**B. Procedure and Instruction**

For both Tiers, we have different procedures designed for the two groups of participants.

Technical Validation (Group 1): The CS students played all five developed subgames (Text Phishing Detective, Password Cracker, Phone Scam AI, SQL Injection Detection, TikTok mini game). Following gameplay, they participated in a structured feedback session where they voted on the most effective game mechanics and provided qualitative feedback regarding technical implementation and engagement flow.

Preference Questionnaire (Group 2): Immediately following the gameplay session, the 50 general participants were interviewed using a Questionnaire designed to measure their attitudes toward gamification and mobile learning. The instrument used a 5-point Likert scale (1=Strongly Disagree, 5=Strongly Agree) to evaluate four key dimensions:

**1) Perceived Learning & Self-Efficacy (Q1, Q2):** Two items measured the educational impact. Participants rated their agreement with statements regarding improved understanding of online security and increased confidence in avoiding risks.

**2) Engagement (Q3, Q4):** Two items assessed user experience, asking participants to rate their attention levels (Scale: Poor to Excellent) and their overall enjoyment (Scale: Strongly Disagree to Strongly Agree).

**3) Format Preference (Q5):** One item specifically evaluated the acceptance of the mobile-first, micro-learning format compared to traditional training methods.

**4) Feature Analysis (Q6, Q7):** Two multi-select questions allowed participants to identify specific elements they found "most memorable" and areas requiring improvement.

### C. Ethical Considerations and Procedure

This study was conducted with formal approval from the Monmouth University Institutional Review Board (IRB). The protocol ensured strict adherence to ethical guidelines:

**1) Electronic Informed Consent:** Before accessing the main game interface, all participants were presented with an electronic consent form. They were required to acknowledge and accept the terms of the study digitally before they could proceed to the gameplay menu.

**2) Anonymity:** Data collection was entirely anonymous; no personally identifiable information (PII) was linked to the performance metrics or survey responses.

**3) Procedure:** The evaluation was conducted in a controlled environment. Participants accessed the five subgames via QR codes displayed on a central screen. Participants were instructed to play through the modules on their personal mobile devices to replicate a realistic "Bring Your Own Device" (BYOD) learning scenario.

### D. Data Analysis

Data from Group 1 (Experts) was analyzed both quantitatively and qualitatively to identify design strengths and technical bugs. Data from Group 2 (General) was analyzed using descriptive statistics (Mean and Standard Deviation) to validate the hypothesis that students prefer mobile, interactive formats over traditional security awareness training.

### E. Comparative Evaluation with Existing Works

Our preliminary work [31] does a comparative evaluation on using interactive (gamified) modules versus Mimecast on phishing detection topics on 51 college participants. Participants were randomly divided into three groups. One trained with gamified modules (18 members), while the other trained with Mimecast (18 members), a widely used commercial awareness platform, the baseline group with no training (15 members).   After 30 minutes of learning, all groups took a detailed online quiz of 25 multiple choice questions on various aspect of email phishing 25 multiple-choice questions designed to assess the effectiveness of the training methods. The result in Table I show that interactive (gamified) learning modules, even though not mobile friendly, has better learning effect on phishing detection topic over traditional platform such as Mimecast to some degree. This evaluation is not the contribution of this paper but helps to clarify the overall effect of gamification compared to existing works.

| Group | Gamified modules | Mimecast | No training |
|---|---|---|---|
| Avg score | 93.33 | 85.33 | 82 |

TABLE I: Evaluation Result of Groups with Training.

## V. RESULTS

The evaluation yielded distinct insights from the technical experts and the general student population, confirming both the quality of the system and the demand for this learning format.

### A. Tier 1: Expert Technical Validation (N=9)

Across all participants, the QR-coded subgames were played collectively more than 50 times. Replay behavior indicated that the Phone Scam and TikTok-style Filter Game were more engaging, with several students choosing to attempt them multiple times. From the voting results, we draw the following findings:

**1) Engagement:** The TikTok mini-Game received the highest engagement rating (7 out of 9 votes), with experts citing its fast-paced nature as highly effective for "attention capture."

**2) Educational Clarity:** The Text Phishing Detective mini-game received 5 votes for "Best Learning Value," as experts noted it effectively taught users to identify URL spoofing and subtle social engineering cues.

**3) Qualitative Feedback:** The experts praised the AI Phone Scam mini-game for its realism. One participant noted that "having an AI chatbot simulate a realistic voice makes the threat feel authentic." However, they also suggested that the Text Scam game needed more balanced answer choices to prevent users from guessing correctly without understanding the underlying concept.

### B. Tier 2: General Learner Preferences (N=50) (Quantitative Analysis)

The survey of 50 general students strongly supported the shift toward mobile-friendly gamification.

**Overall Analysis:** When it comes to the confidence and learning topic (Q1, Q2), self-reported efficacy was high. 90% of students (45/50) agreed or strongly agreed that they felt more confident in avoiding security risks after playing the game. Similarly, 78% (39/50) confirmed the game helped them better understand how to stay secure online. For enjoyment and attention (Q3 & Q4), 88% of participants reported enjoying the game (Q4), and 78% stated that it successfully kept their attention (Q3). Additionally, engagement metrics were strong, with 44 users reporting they enjoyed the game (Q4) and 39 users stating it successfully kept their attention (Q3).

The aggregated survey results demonstrate a strong validation of the mobile-first gamification approach. In question 5 about format preference (Q5), the most significant finding was the overwhelming preference for the delivery method. When asked if they preferred this type of short, mobile-friendly game over long desktop training or videos, 92% of participants (46/50) agreed or strongly agreed. Notably, 72% (36/50) selected "Strongly Agree," indicating a decisive shift in student preference toward micro-learning formats.

**Feature Impact Analysis:** Participants were asked to vote on which specific elements contributed most to their positive experience, and the aggregate data strongly highlights the value of "bite-sized" content. As shown in Figure 8(left), The "Short Length / Quick Format" received the highest number of votes (23 votes), a finding that correlates directly with the strong preference for micro-learning observed in Question 5. Following closely was the "Ease of Learning," which garnered 18 votes. Additionally, the "Phone Scam Simulation" received 16 votes, reinforcing the qualitative feedback that the audio realism provided a standout and authentic training experience.

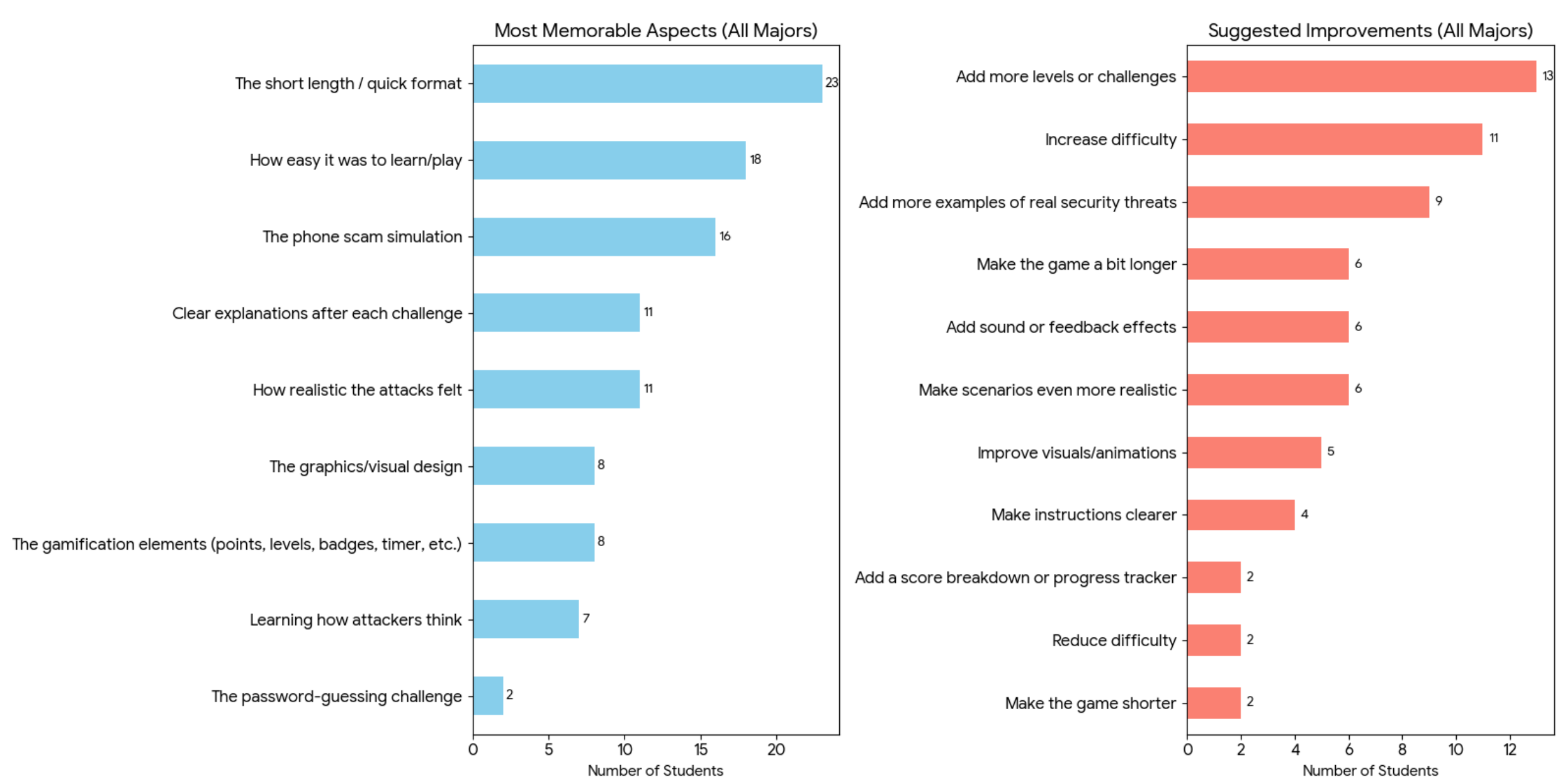

Fig. 8: The results of Feature Impact(Q6) and Areas for Improvement(Q7) in Survey.

**Areas for Improvement:** Quantitative data regarding desired improvements points toward a need for content expansion rather than structural changes in Figure 8(right). The top request was for "Content Expansion," with 13 participants voting to add more levels or challenges. "Difficulty Scaling" was also a priority, with 11 participants voting to increase difficulty; this request was significantly higher among the female cohort (8 votes) compared to the male cohort (3 votes), suggesting future iterations should employ adaptive difficulty. Finally, 9 participants requested "More examples of real security threats," indicating a desire for a broader variety of attack vectors to enhance realism.

**Comparative Analysis by Gender:** While the overall reception was positive, a breakdown of the data reveals distinct nuances in how male and female participants engaged with the gamified elements. The average scores of questions in the survey are shown in Figure 9.

***Impact on Confidence:*** The gamified approach appeared particularly effective for the female cohort. In response to Q2, 100% of female participants (25/25) responded positively. In contrast, while the male cohort was largely positive, a small subset (8%, n=2) expressed strong disagreement.

***Format Acceptance:*** Both groups favored the mobile format, but the intensity of this preference varied. Female participants exhibited a stronger preference, with 88% (22/25) selecting "Strongly Agree" for Q5. Male Participants: While supportive, only 56% (14/25) selected "Strongly Agree," with a larger portion selecting "Agree." This suggests that the shift to mobile micro-learning is a critical requirement for engaging female students, while male students remain slightly more tolerant of diverse formats.

The quantitative data supports the conclusion that the mobile-friendly gamification model is highly effective, particularly in building security confidence among female students and non-technical majors. The disparity in feedback suggests that future iterations should focus on two parallel tracks: enhancing visual fidelity to appeal to male users, and implementing adaptive difficulty levels to satisfy the female users' desire for greater challenges.

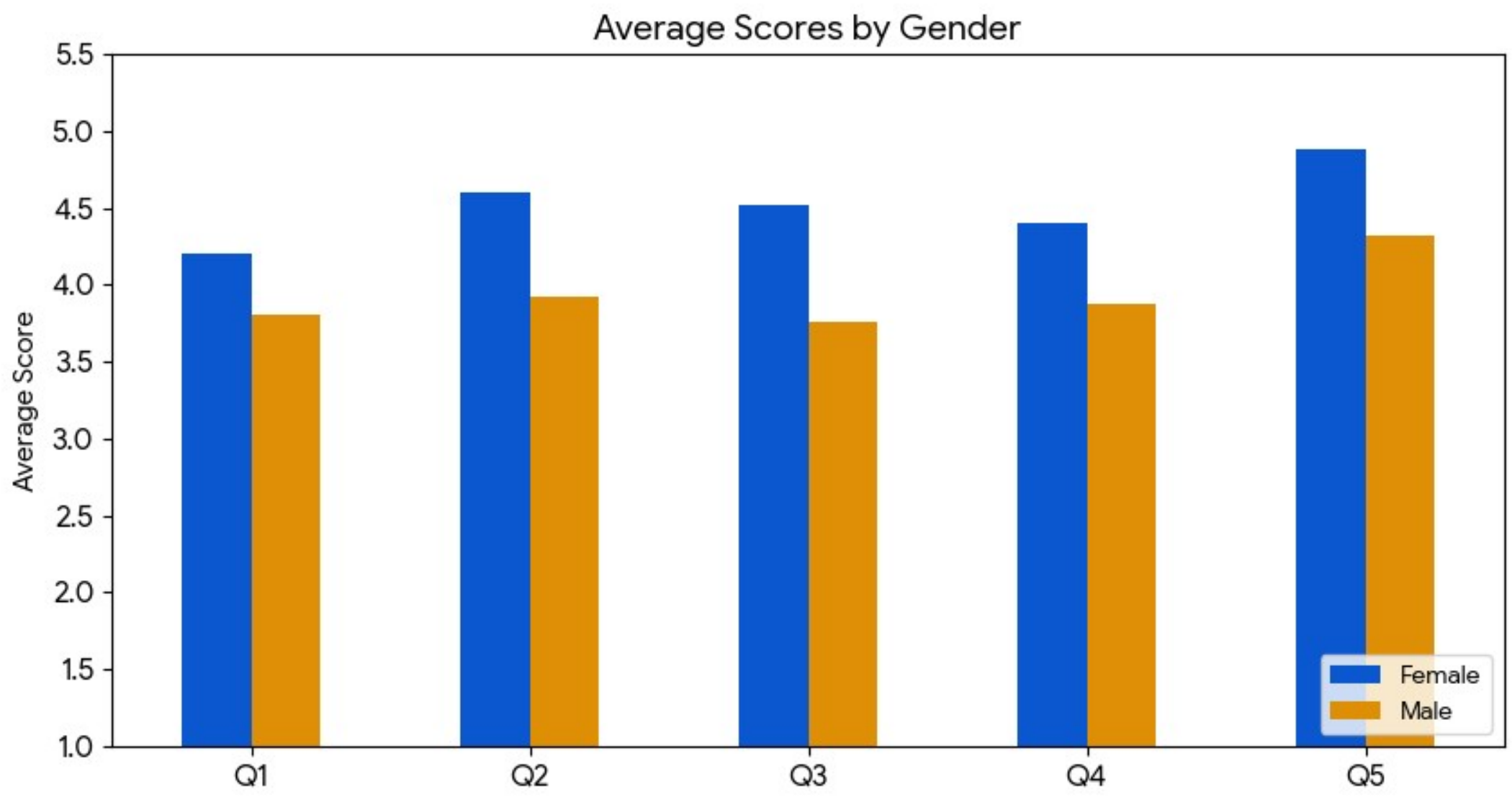

Fig. 9: Average Scores by Gender in Survey.

**C. Tier 2: General Learner Preferences (N=50) (Qualitative Analysis)**

In addition to the quantitative results, participants also provided rich open-ended feedback through a "Brainstorming Section." This qualitative data was analyzed to identify recurring themes regarding learning retention, game mechanics, and future content direction.

**Learning Retention (What Stuck?):** When asked what they remembered most from the training, participants consistently cited specific interactive elements and key takeaways, indicating that the gamified approach successfully anchored memory. The "Password Decoding Game" and "Guessing the Phishing Emails" were frequently highlighted as the most memorable components, suggesting that active participation leads to better recall. Beyond the mechanics, users retained critical vulnerability awareness, specifically noting that elderly people are most likely to fall for fake or unsafe emails, while teens and college students are also high-risk targets. Participants also internalized actionable verification habits, such as the need to "always double check who the sender is" and to actively look for "strange patterns" in URLs before clicking.

**Critical Feedback on Mechanics:** Despite the positive reception, participants provided constructive criticism regarding the current game design, specifically focusing on the rigor of the assessments. A dominant theme in the feedback was the predictability of the answer choices; multiple participants noted that "Most answers are 'all of the above,'" which allowed them to guess correctly without fully processing the underlying educational content. To remedy this, users suggested that future iterations should "make the answers more randomized" to eliminate these obvious patterns. Furthermore, there was a strong desire for greater interaction depth. Participants argued for active testing methods, such as typing in answers rather than simple selection, stating, "We won't really learn that way" if the experience remains passive. Feedback also pointed toward difficulty scaling, with suggestions to implement higher stakes, such as a "Baseball" style rule where three wrong answers result in a loss, or to replace time limits with point deductions to balance challenge with anxiety reduction.

**User Segmentation and Personalization:** The most prolific feedback concerned the need to tailor the experience to specific demographics, rather than employing a "one size fits all" approach. Participants strongly recommended segmenting the training content based on age groups to maximize relevance. For children (ages 4–10), suggestions favored simple, game-based interactions. For teens and college students, participants requested topics relevant to their daily digital lives, such as safety on dating apps ("Catfishing games"), "Fortnite/Madden" themed scenarios, and detecting AI-generated media. Conversely, feedback emphasized that training for the elderly (50+) should focus on scenarios like coupons, shopping deals, and lottery wins to "resonate with the everyday emails" they actually receive. A key insight from this feedback was the proposal for an "Enter Age" feature at login, which would automatically filter and deliver the most appropriate content for the user's demographic profile.

**Feature Requests and Future Innovation:** Participants brainstormed an extensive list of features to make security awareness more interesting and modern. There was a significant demand for content covering emerging threat vectors, specifically AI deepfakes, AI voice cloning, and catfishing, reflecting a desire to stay ahead of sophisticated social engineering attacks. In terms of gamification, requests included the addition of leaderboards, multiplayer or team competitions, and profile customization to enhance long-term engagement. Users also proposed creative new game modes, such as a "Hangman" style password cracking game, "Candy Crush" inspired puzzles, and seasonal themes (e.g., winter-themed challenges) to keep the experience fresh and visually appealing.

**Conclusion of Qualitative Findings:** The qualitative data confirms that while the concept of gamified training is highly engaging, the execution needs to evolve from passive recognition to active recall. The reliance on predictable multiple-choice patterns was identified as a barrier to deep learning, prompting a need for randomized answers and more rigorous testing methods. Furthermore, the feedback strongly supports a shift toward a segmented model, where age-relevant scenarios, ranging from dating app safety for students to financial scams for the elderly, ensure that the training maximizes personal relevance and real-world applicability.

## VI. DISCUSSION

This study illuminates the transformative potential of AI-powered gamification in bridging the persistent gap between theoretical cybersecurity knowledge and practical behavioral application. Our findings suggest that the efficacy of the Sentinel platform extends beyond simple engagement; it fundamentally alters the cognitive approach of the learner. By shifting the pedagogical focus from passive compliance to active resilience, the platform addresses specific psychological nuances that traditional video-based training often overlooks.

The distinct success of the "Password Cracker" and "TikTok-style" modules suggests that gamified mechanics are most effective when they compel learners to adopt an offensive, rather than purely defensive, mindset. While traditional training encourages students to memorize rules, the "Password Cracker" mini-game requires them to deduce credentials based on a fictional target's background. This mechanic fosters adversarial thinking, forcing students to internalize the logic of an attacker and actively assess data vulnerabilities rather than merely following a checklist. Similarly, the imposition of time constraints in the TikTok filter game does more than create urgency; it mimics the cognitive load of real-world mobile usage. As participants noted that this format effectively raised awareness in under 30 seconds, it is evident that aligning educational delivery with the fast-paced, "System 1" decision-making habits of digital natives can significantly lower the barrier to entry for complex security concepts. Furthermore, the integration of Generative AI distinguishes this platform from static role-playing exercises by introducing dynamic behavioral conditioning. Participants explicitly emphasized that the "realistic voice" and responsiveness of the AI phone scammer created a sense of authenticity absent in standard quizzes. This feedback indicates that AI serves a crucial function in simulating the emotional and psychological pressures of social engineering. By engaging with an adaptive agent that mimics manipulative tactics, learners practice emotional regulation, such as the ability to pause and verify under stress, a skill that cannot be acquired through passive information consumption.

## VII. ETHICAL CONSIDERATION AND AI GOVERNANCE

While AI-powered gamification offers significant benefits, it introduces specific ethical risks that must be managed, particularly in an educational context.

**Model Hallucination and Accuracy:** There is an inherent risk that a generative AI model might provide inaccurate information. To mitigate this, we strictly limit the scope of the generative model to the adversarial role only. The AI is exclusively employed to simulate the attacker (e.g., generating phishing text or scam calls), where minor inconsistencies do not harm the learning objective. The educational component, explaining why an attack works or how to prevent it—is delivered through standard, non-generative instructional interfaces to guarantee 100\% pedagogical accuracy.

**Bias and Stereotyping:** AI models can inadvertently perpetuate biases, such as associating specific accents or dialects with criminal behavior. To address this, the Phone Scam module utilizes a randomized set of voice profiles (varying in gender, tone, and accent) to prevent reinforcing stereotypes about the demographics of cyber attackers.

**Privacy and Data Leakage:** Given that the system processes voice and text inputs, user privacy is paramount. Audio data from the phone scam game is processed ephemerally; voice inputs are transcribed to text for the model and immediately discarded, not stored. At the same time, as detailed in the methodology, no PII is associated with the interaction logs. The system creates a temporary session ID for the duration of the game which is purged upon completion.

This study was conducted in accordance with the journal’s Publication Ethics and Malpractice Policy. All participants were informed of the study’s purpose, and participation was voluntary. No personally identifiable information was collected, and all data were handled in compliance with institutional and ethical guidelines for research involving human subjects.

## VIII. CONCLUSION

Cybersecurity education in higher education faces ongoing challenges as digital threats grow more sophisticated. Traditional awareness programs often rely on static content that does not fully engage learners or prepare them for the complexities of modern attacks. The real-world phishing incident discussed in this paper illustrates how AI-driven scams increasingly exploit human behavior in ways that conventional training cannot effectively address.

This study explored how gamification, enhanced by artificial intelligence(AI), can provide a more engaging and adaptive approach to cybersecurity education. By leveraging Generative AI to create responsive, realistic scenarios (such as the "Phone Scam" simulator) and delivering them through mobile-friendly mini-games, we achieved high engagement and learning outcomes. Our two-tiered evaluation confirmed that technical experts validated the realism of the simulations, while the general student population ($N=50$) overwhelmingly preferred this mobile-first approach over traditional desktop training.

In the future research, we plan continue to refine adaptive gamified systems and examine their effectiveness across different student populations and learning environments. And we also identify three critical directions to advance this research:

1) Curricular Integration and Forensic Expansion: While the current platform operates as a standalone awareness tool, future work will focus on integrating these modules into university Learning Management Systems (LMS).

2) Policy Implications for Higher Education: This research suggests a necessary shift in institutional policy from "Compliance-Based" to "Resilience-Based" metrics. Universities currently measure success by the percentage of students who watch a video such as Mimecast. We advocate for policies that leverage gamified data to measure resilience, tracking how students perform in simulated attacks over time.

3) Addressing AI-Enabled Threats: since attackers increasingly use deepfakes and voice cloning, education tools must mirror these capabilities. Future iterations of the platform will explore the use of deepfake-detection mini-games, training users to identify visual and auditory artifacts in AI-generated media.

## IX. CONFLICT OF INTEREST

The authors declare that they have no conflicts of interest or competing financial or personal relationships that could have influenced the work reported in this paper.